\documentclass[sort&compress,numrefs]{wiley-article}

\usepackage{graphicx}
\usepackage[space]{grffile}
\usepackage{latexsym}
\usepackage{textcomp}
\usepackage{longtable}
\usepackage{tabulary}
\usepackage{booktabs,array,multirow}
\usepackage{amsfonts,amsmath,amssymb}
\usepackage{natbib}
\usepackage{url}
\usepackage{hyperref}
\hypersetup{colorlinks=false,pdfborder={0 0 0}}
\usepackage{etoolbox}

\newif\iflatexml\latexmlfalse
\AtBeginDocument{\DeclareGraphicsExtensions{.pdf,.PDF,.eps,.EPS,.png,.PNG,.tif,.TIF,.jpg,.JPG,.jpeg,.JPEG}}

\usepackage[utf8]{inputenc}
\usepackage[english]{babel}

\usepackage{siunitx}
\usepackage{placeins}
\usepackage{braket}
\usepackage{simplewick}
\usepackage{pdflscape}

\newcommand{\PPM}[0]{\mathcal{P}}
\newcommand{\LI}[0]{\mathcal{L}_{I}}
\newcommand{\LM}[0]{\mathcal{L}_{M}}
\newcommand{\LP}[0]{\mathcal{L}_{P}}
\newcommand{\LQ}[0]{\mathcal{L}_{Q}}

\newcommand{\MV}[0]{\widetilde{\Psi}}
\newcommand{\HINT}[0]{\widetilde{H}}
\newcommand{\VINT}[0]{\widetilde{V}}

\iflatexml

\else

\paperfield{Field of the paper}
\corraddress{
Andr\'ei Zaitsevskii, National Research Center ``Kurchatov Institute'', B.P. Konstantinov Petersburg Nuclear Physics Institute, Gatchina, Leningrad District 188300, Russia
}
\corremail{zaitsevskii\_av@pnpi.nrcki.ru}
\fundinginfo{Russian Science Foundation, \\ Grant/Award Number: 20-13-00225; \\
Ministry of Science and Higher Education of the Russian Federation, \\ Grant/Award Number: 075-10-2020-117.
}
\fi

\papertype{Original Article}

\title{Generalized relativistic small-core pseudopotentials accounting for quantum electrodynamic effects: construction and pilot applications}

\author[1,2]{Andr\'ei Zaitsevskii}
\author[1]{Nikolai S. Mosyagin}
\author[1]{Alexander V. Oleynichenko}
\author[3]{Ephraim Eliav}

\affil[1]{Petersburg Nuclear Physics Institute named by B.P.\ Konstantinov of National Research Center ``Kurchatov Institute'' (NRC ``Kurchatov Institute'' - PNPI), 1 Orlova roscha, Gatchina, 188300 Leningrad region, Russia}

\affil[2]{Department of Chemistry, M.V.~Lomonosov Moscow State University, Leninskie gory 1/3, Moscow, 119991~Russia}

\affil[3]{School of Chemistry, Tel Aviv University, 6997801 Tel Aviv, Israel}

\runningauthor{Andr\'ei Zaitsevskii}

\begin{document}

\maketitle
\selectlanguage{english}
\begin{abstract}
A simple procedure to incorporate one-loop quantum electrodynamic (QED) corrections into the generalized (Gatchina) nonlocal shape-consistent relativistic pseudopotential model 
is described. The pseudopotentials for Lu, Tl, and Ra replacing only inner core shells (with principal quantum numbers $n\le 3$ for the two former elements and $n\le 4$ for the latter one) are derived from the solutions of reference atomic SCF problems with the Dirac-Coulomb-Breit Hamiltonian to which the model Lamb shift operator added. QED contributions to atomic valence excitation energies evaluated at the SCF level are demonstrated to exceed the errors introduced by the pseudopotential approximation itself by an order of magnitude. Pilot applications of the new model to calculations of excitation energies of two-valence-electron atomic systems using the inter\-me\-diate-Ha\-mil\-to\-nian relativistic Fock space coupled cluster method reformulated here for incomplete main model spaces are reported. Implications for high-accuracy molecular excited state calculations are discussed.

\end{abstract}

\section{Introduction}

Accurate
\emph{ab initio} electronic structure modeling on heavy-element compounds implies going beyond non-relativistic quantum mechanics. Since the 
1990s, atomic and molecular electronic structure calculations with Dirac-Coulomb Hamiltonian were made routinely available~\cite{Schwerdtfeger:04,Barysz:10,Liu:16}; however, the insufficient accuracy of this Hamiltonian for numerous purposes, especially for excited state studies and simulation of high-resolution spectra, has gradually become apparent. For instance, contributions of two-electron frequency-independent Breit interactions to electronic excitation energies can reach several hundreds of wavenumbers even for the 6th row elements \cite{Eliav:96,Kahl:21}, being significantly more pronounced for actinides~\cite{Infante:07,Safronova:11,Eliav:15,Mosyagin:16}. These contributions are especially 
sizeable for processes accompanied by the change of the occupation number of the $5f$ shell in actinides.
Atomic calculations accounting for the Breit operator 
have been available since the 1990s \cite{Eliav:94}, whereas the lack of general purpose codes for evaluation and transformation of Breit integrals interfaced to the most commonly used program packages for \emph{ab initio} molecular electronic structure modeling restricts the ability of molecular studies with full Dirac-Coulomb-Breit Hamiltonian (see, however, Refs.~\cite{Maison:20,Skripnikov:21,Skripnikov:QED:21} for recent implementations and pilot applications). At the same time, the relativistic pseudopotential (RPP) approach, and in particular its generalized (Gatchina) version, GRPP (the most widely used form of generalized relativistic effective core potential, GRECP),  provides an attractive opportunity to efficiently account for the bulk of Breit interactions in a very economical way \cite{Petrov:04b,
Mosyagin:06,Mosyagin:10,Mosyagin:16,Mosyagin:20}. 
High accuracy of GRPP in reproducing the effect of Breit interactions in molecular 
property calculations was demonstrated recently~\cite{Mosyagin:21,Zaitsevskii:22}. An alternative technique of simulating the instantaneous magnetic (Gaunt) part of Breit interactions based on the molecular mean field procedure \cite{Sikkema:09} 
has been implemented on the correlation step
in the DIRAC program suite~\cite{DIRAC_code:19,DIRAC:20}; it is to be noted that its use implies markedly more time-consuming computations.

Furthermore, the magnitude of one-loop quantum-electrodynamics (QED) contributions 
(electron self energy and vacuum polarization) to 
atomic and molecular 
properties can be comparable with the 
Breit/Gaunt one and even exceed 
it~\cite{Thierfelder:2010,Sunaga:QED:22}. 
A systematic QED treatment is currently available only for the few-electron atoms~\cite{Shabaev:02,Shabaev:06}. Apparently, the simplest way to account for the lowest-order QED effects is to use the model Lamb shift operator technique~\cite{Flambaum:05,Shabaev:13,Shabaev:18,Malyshev:22}. In recent years it became routinely exploited in atomic calculations \cite{Thierfelder:2010,Schwerdtfeger:15,%
Pasteka:17,Oleynichenko:CCSDT:20,Kahl:21,Guo:21,Kaygorodov:21,Kaygorodov:22}; 
successful adaptations to molecular studies within the four-component methodology were presented in \cite{Sunaga:21,Skripnikov:QED:21,Sunaga:QED:22}. In the recent paper~\cite{Sunaga:QED:22}, the very high importance of QED effects for accurate calculations of spectroscopic and even thermodynamical properties of heavy-element-containing molecules was clearly demonstrated. Thus incorporation of QED effects directly into the relativistic pseudopotential model seems to be a logical step toward the cheap and versatile account of these effects in molecular calculations (including those for rather complex systems). The first semilocal RPP model with QED contributions was constructed by Hangele and coauthors \cite{Hangele:12, Hangele:13} using the ``energy adjustment'' methodology and thus explicitly taking into account only the QED effect on the reference atomic energy spectra. The level of their pseudopotential errors was comparable with the QED corrections themselves, apparently due to the relatively small number of explicitly treated electrons and limitations of  the semilocal form of RPPs.

In the present paper we follow a different strategy, reproducing 
the QED effects on the wavefunctions directly \emph{via} imposing the shape-consistency requirement~\cite{Lee:77,Hafner:79,Titov:99}. The use of flexible GRPP Ansatz provides an opportunity to increase the number of explicitly treated electronic shells without losing accuracy in describing any of these shells, thus reducing the errors in estimating the core-valence contributions to various molecular properties in correlation calculations. 

From a practical standpoint, the high-precision basic electronic structure model defined by the new GRPPs can be useful only if the errors introduced by approximate correlation treatment 
is lower than 
or at least comparable to the magnitudes of QED corrections.   
The Fock space multireference coupled cluster method~\cite{Lindgren:78,Kaldor:91,Bartlett:07,Lyakh:11,Musial:11} has proven itself as one of the most prospective tools for highly accurate predictions of atomic and molecular spectra, properties of atoms and molecules in their ground and excited electronic states and transition moments between these states~\cite{Bhattacharya:13,Zaitsevskii:TDM:18,Zaitsevskii:20}. This method can be readily used in conjunction with various relativistic approximations to the many-electron Hamiltonian (the relativistic Fock space coupled cluster theory, FS-RCC) \cite{Eliav:98,Visscher:01}, including those involving QED terms~\cite{Eliav:15b,Eliav:17,Eliav:Review:22}. 
The most widely used FS-RCC model which accounts for only single and double excitations in the cluster operator (FS-RCCSD) generally cannot be regarded as complete enough \cite{Oleynichenko:CCSDT:20,Basumallick:21,Porsev:21,Skripnikov:21} (see also \cite{Haque:85,Hughes:93,Musial:08,Musial:19} for the non-relativistic case). The error arising from the neglect of higher excitations varies in a wide range for different systems; in general, the role of triples is especially important for electronic states with more than one valence quasiparticle. With rather rare exceptions (e.~g., superheavy elements), the expected FS-RCCSD error for electron excitation energies exceeds one due to the neglect of QED effects. Full FS-RCCSDT calculations involving non-perturbative evaluation of the amplitudes of triple excitations are extremely cumbersome and currently can be performed with very restricted one-electron bases (a few hundreds of spinors), whereas perturbative estimates, in contrast with the single-reference CCSD(T) scheme, are surprisingly inaccurate~\cite{Vaval:98,Bernholdt:99,Meissner:04}. Only a part of triples can be covered simply by a model space extension (see \cite{Oleynichenko:CCSDT:20,Skripnikov:21} for the detailed discussion). A reasonable way to achieve sufficient accuracy in accounting for triple excitations might consist in resorting to additive schemes \cite{Oleynichenko:CCSDT:20,Skripnikov:QED:21,Zaitsevskii:22}.

Among the weaknesses of the Fock space coupled cluster theory, the intruder state problem \cite{Evangelisti:87} must be mentioned. It is inherent for multireference coupled cluster approaches operating with complete (or quasicomplete) model spaces and manifest itself as the divergence of Jacobi scheme of solving amplitude equations (see, for example, \cite{Zaitsevskii:18a,Eliav:Review:22} and references therein). As for Fock space formulations, the intruder state problem is quite common for the sectors with two or more quasiparticles over the closed shell vacuum. A vast variety of solutions 
for the problem were proposed in several decades, including the use of incomplete model spaces \cite{BenShlomo:88,Kaldor:88}, conversion of the 
amplitude equations into matrix eigenvalue problems (equation-of-motion (EOM)-like formulations, \cite{Meissner:95,Meissner:96,Meissner:98,Musial:08,Meissner:10}), construction of intermediate Hamiltonians~\cite{Landau:00,Landau:01,Eliav:05}, shifts of energy denominators with subsequent extrapolation to the zero-shift limit \cite{Eliav:05,Zaitsevskii:18a} etc. None of these approaches can be regarded as the perfect one. For instance, the EOM-like formulations imply the construction and diagonalization of huge matrices; this can be very computationally demanding and thus prohibitive for either relativistic treatment of large systems or incorporation of triple (or higher) excitations into the cluster operator. Being numerically stable, such schemes do not provide a general mean to combine the continuity of solutions as functions of molecular geometry with moderate amplitude values in wide ranges of geometries. The accuracy of approaches based on the energy denominator shifting or complete suppression of problematic amplitudes \cite{Zaitsevskii:17,Zaitsevskii:18a,Oleynichenko:20cpl,Kruzins:21,Eliav:Review:22} can hardly be estimated and may become insufficient in the studies beyond the FS-RCCSD level. The convergence of results with respect to different values of the shifting parameters have to be thoroughly studied in each specific case. The values of the parameters themselves have to be chosen manually, which definitely does not make this method easy-to-use nor user friendly. To sum up, it can be argued that a fresh look 
at the intruder state problem is required. The need for manual parameter setting should be minimized, and possible increases of errors should be efficiently diagnosed without any reference to external data.

The paper is organized as follows. In Section~\ref{sec:qedrecp}, the procedure of 
incorporating
QED effects into shape-consistent relativistic pseudopotentials is described and applied to construct the GRPP for Lu, Tl,
and Ra.
The next section presents the formulation of the intermediate-Hamiltonian FS-RCC method with incomplete main model space which is used in pilot calculations of 
two-valence atomic systems in Section~\ref{sec:results}. The final section discusses the prospect of molecular applications of the new approach and provides some conclusive remarks.

\section{Theoretical considerations}

\subsection{Shape-consistent relativistic pseudopotentials accounting for QED effects}
\label{sec:qedrecp}

The generalized (Gatchina) 
pseudopotential (GRPP) method has proven itself as very accurate one in many calculations (see the latest reviews~\cite{Mosyagin:16, Mosyagin:17, Mosyagin:20}). To include QED effects into the GRPP operator, the generation scheme~\cite{Tupitsyn:95, Mosyagin:97, Petrov:04b, Mosyagin:06} 
(which, in turn, is based on ``shape-consistent'' semilocal RPP generation scheme~\cite{Christiansen:79,Goddard:68,Lee:77,Ermler:78,Hafner:79,Pacios:85}) was applied with the single modification: the HFD-QED code~\cite{Shabaev:13,Shabaev:18} recently developed by the group from Saint-Petersburg University (Russia) was used at the first step instead of the HFD code~\cite{Bratzev:77}. In the framework of the HFD-QED code, the one-electron vacuum polarization operator is given by the sum of the local Uehling and Wichmann-Kroll potentials~\cite{Mohr:98b}, whereas the self energy operator is modeled with the help of some semilocal and nonlocal GRPP-like operators (see Eqs.~(13) and~(15) in Ref.~\cite{Shabaev:13}). Thus, there are no theoretical problems to include the above mentioned QED effects 
directly into the atomic GRPP operator~\cite{Titov:99} which consists of the local, semilocal and nonlocal parts:
\begin{eqnarray}
  \hat{U}^{\rm GRPP} & = & U_{n_vLJ}(r) + \sum_{l=0}^L \sum_{j=|l-1/2|}^{l+1/2} 
                   \bigl[U_{n_vlj}(r)-U_{n_vLJ}(r)\bigr] 
                   \hat{P}_{lj}                             \nonumber\\ 
             & + & \sum_{l=0}^L \sum_{j=|l-1/2|}^{l+1/2} \sum_{n_c} 
                   \Bigl\{\bigl[U_{n_clj}(r)-U_{n_vlj}(r)\bigr] 
                   \hat{P}_{n_clj} + \hat{P}_{n_clj}  
                   \bigl[U_{n_clj}(r)-U_{n_vlj}(r)\bigr]\Bigr\} \nonumber\\ 
             & - & \sum_{l=0}^L \sum_{j=|l-1/2|}^{l+1/2} \sum_{n_c,n_c'} 
                   \hat{P}_{n_clj} 
                   \biggl[\frac{U_{n_clj}(r)+U_{n_c'lj}(r)}{2} - U_{n_vlj}(r)\biggr] 
                   \hat{P}_{n_c'lj}
.
\label{GRPP_jj1}
\end{eqnarray}
Here $r$ denotes the distance from the center of the atomic nucleus, $\hat{P}_{lj}$ projects onto the subspace of one-electron two-component functions with spatial  angular momentum $l$ and total  angular momentum $j$ with respect to this nucleus, $\hat{P}_{n_clj}$ stands for the projector onto the subspace of subvalence (outercore) pseudospinors with 
the principal quantum number $n_c$ and angular momenta $l$ and $j$, $n_v$ is the principal quantum number of valence electron. The functions $U_{nlj}(r)$ are radial GRPP components (partial potentials); $L$ is normally one more than the highest orbital angular momentum $l_{max}$ of the excluded from GRPP calculations (innercore) spinors, and $J=L+1/2$.

Thus, the many-electron Dirac-Coulomb-Breit (DCB) Hamiltonian with Uehling, Wichmann-Kroll, model self-energy, and Fermi nuclear charge distribution 
potentials is replaced by the effective Hamiltonian with the GRPP
\begin{equation}  
H = \sum_{q} \left[- \frac{1}{2} {\bm{\nabla}}_q^2 - 
\sum_\gamma
\frac{Z^*_\gamma}{r_{\gamma q}} + 
\sum_\gamma \hat{U}^{\rm GRPP}_\gamma(\mathbf{r}_q - \mathbf{r}_\gamma
,\sigma_q)\right] + \sum_{q > q'} \frac{1}{r_{qq'}}\label{Hef}
\end{equation}
where $(\mathbf{r}_q,\ \sigma_q)$ denote the set of spatial and spin coordinates of the $q^{\rm th}$ electron, $\mathbf{r}_\gamma$ 
stand for the set of spatial coordinates of the nucleus $\gamma$, $r_{\gamma q}$ and $r_{qq'}$ are the distances between the nucleus $\gamma$ and the $q^{\rm th}$ electron and between the electrons $q$ and $q'$, respectively, 
$Z_\gamma^*$ is the effective innercore charge of the atom $\gamma$ (its nuclear charge $Z_\gamma$ minus the number $N_{\gamma i}$ of  innercore electrons); the summations run over the indices of valence and subvalence (outercore) electrons and atomic nuclei. The GRPP operator in Eq.~(\ref{Hef}) is a superposition of atomic contributions~(\ref{GRPP_jj1}). It should be noted that some 
atoms $\gamma_a$ can be described by the conventional all-electron nonrelativistic way (i.~e.\ without GRPP), then it is obvious that $\hat{U}^{\rm GRPP}_{\gamma_a}=0$, $N_{{\gamma_a} i}=0$, and $Z_{\gamma_a}^*=Z_{\gamma_a}$.
In contrast to the four-component wave function used in DC(B) calculations, 
one-electron pseudo-wave functions in the case of full or scalar relativistic calculations with the GRPP is two- or one-component.

To make the incorporation of QED effects into the GRPP operator practically valuable, the level of GRPP errors should be essentially lower than
the size of the QED contributions. The required accuracy is usually attained in the case of the GRPPs with {\it tiny} cores when valence and {\it several} outercore shells with the same quantum numbers $lj$ are explicitly treated in the GRPP calculations. In the present paper, the new GRPP versions were generated for the Ra and Tl, Lu atoms with 60 and 28 respectively innercore electrons excluded from the following calculations with these GRPPs. The GRPP components were constructed for the 
$5spdfg,6spd,7sp$ shells of Ra and the $4spdf,5spd,6sp$ shells of Tl, Lu
(where $6sp$ mean $6s_{1/2}$, $6p_{1/2}$, $6p_{3/2}$ and similarly for others). 
Then the valence and core GRPP versions were derived from the above (full) GRPP versions by neglecting the differences between the outercore and valence potentials. Thus, the valence GRPP operators for the Ra or Tl, Lu atoms are semilocal ones with the $7sp,6d,5fg$ or $6sp,5d,4f$ components of the full GRPP version (i.~e.\ the first line in Eq.~(\ref{GRPP_jj1})). The main difference between the valence GRPPs and the conventional shape-consistent
RPPs is that the components of the former are deduced for nodal valence pseudospinors. In the current GRPP versions, the $s$ and $p$ valence pseudospinors have two nodes, whereas the $d$ ones have a single node. Thus, these are the valence
potentials (not the outercore or somehow averaged ones) that act on the valence electrons in this GRPP version. The value of different contributions into all-electron Hamiltonian as well as the accuracy of the different GRPP versions 
are demonstrated in Tables~\ref{tab:Ra-GRPP}, \ref{tab:Tl-GRPP}, and \ref{tab:Lu-GRPP} with the help of the numerical self-consistent field (SCF) calculations on the Ra, Tl, Lu atoms and its cations. One can see that 
a proper account for the finite nuclear size, Breit interactions, and QED effects 
is essential for achieving high accuracy.
In particular, for Tl, the maximal errors of the
full and valence GRPP versions are 17 and 67 cm$^{-1}$ respectively, whereas the maximal contributions of the QED effects, finite nuclear size, and Breit interactions are 543, 198, and 237 cm$^{-1}$ among all the possible transitions between the states considered in Table~\ref{tab:Tl-GRPP}.

\begin{table}[h]
\caption{Excitation energies derived from all-electron 
numerical SCF calculations for the states averaged over nonrelativistic configurations of the Ra atom and its cations with DCB Hamiltonian and accounting for the finite nuclear size and QED effects.
The contributions from various effects are the differences from 
the results of the all-electron numerical SCF calculations without the QED effects, with the point nucleus, or without Breit interactions. Errors of the GRPPs are estimated as differences between the results of the all-electron calculations and the calculations with corresponding GRPP versions. All data are in cm$^{-1}$.}
\begin{tabular}{lrrrrrrrr}
\hline
& \multicolumn{1}{c}{\textbf{Excitation energy}} & & \multicolumn{3}{c}{\textbf{Contributions from}} & & \multicolumn{2}{c}{\textbf{Errors of}} \\
      \cmidrule{2-2} \cmidrule{4-6} \cmidrule{8-9}
$\ldots 6s^2 6p^6 7s^2\; \rightarrow$ 
                     & \textbf{DCB+QED, finite nucl.}    & & \textbf{QED} & \textbf{Finite nucl.} & \textbf{Breit} & & \textbf{full GRPP}  & \textbf{valence GRPP} \\
 \hline
 $\ldots 6s^2 6p^6 7s^1 7p^1$ &        12502 & &  $-$56 &    $-$34 &     4 & &   0 &   3 \\
 $\ldots 6s^2 6p^6 7s^1 6d^1$ &        13445 & &  $-$87 &    $-$46 &   $-$68 & &  $-$3 &  $-$9 \\
 $\ldots 6s^2 6p^6 7s^1     $ &        35099 & &  $-$43 &    $-$26 &   $-$10 & &   0 &   7 \\
 $\ldots 6s^2 6p^6 6d^1     $ &        48499 & & $-$157 &    $-$85 &  $-$102 & &  $-$3 &  $-$9 \\
 $\ldots 6s^2 6p^6 7p^1     $ &        56993 & & $-$125 &    $-$76 &    $-$5 & &   3 &  16 \\
 $\ldots 6s^2 6p^6          $ &       111215 & & $-$125 &    $-$75 &   $-$35 & &   3 &  26 \\
\hline
\end{tabular}
\label{tab:Ra-GRPP}
\end{table}

\begin{table}[h]
\caption{SCF excitation energies for Tl. See the caption of Table~\ref{tab:Ra-GRPP}.}
\begin{tabular}{lrrrrrrrr}
\hline
& \multicolumn{1}{c}{\textbf{Excitation energy}} & & \multicolumn{3}{c}{\textbf{Contributions from}} & & \multicolumn{2}{c}{\textbf{Errors of}} \\
      \cmidrule{2-2} \cmidrule{4-6} \cmidrule{8-9}
$\ldots 5d^{10} 6s^2 6p^1\; \rightarrow$ 
                     & \textbf{DCB+QED, finite nucl.}    & & \textbf{QED} & \textbf{Finite nucl.} & \textbf{Breit} & & \textbf{full GRPP}  & \textbf{valence GRPP} \\
 \hline
 $\ldots 5d^{10} 6s^2     $ &        40 075 & &     35 &       16 &   $-$64 & &    2 &  $-$2 \\
 $\ldots 5d^{10} 6s^1 6p^2$ &        50 829 & & $-$230 &    $-$85 &   $-$35 & &    5 & $-$25 \\
  $\ldots 5d^{10} 6s^1 6p^1$ &        93 066 & & $-$218 &    $-$78 &   $-$92 & &    6 & $-$30 \\
$\ldots 5d^{10} 6p^3     $ &       116 815 & & $-$485 &   $-$179 &   $-$76 & &   17 & $-$45 \\
 $\ldots 5d^{10} 6p^2     $ &       161 289 & & $-$492 &   $-$180 &  $-$129 & &   17 & $-$53 \\
 $\ldots 5d^{10} 6s^1     $ &       188 267 & & $-$210 &    $-$70 &  $-$208 & &    7 & $-$41 \\
 $\ldots 5d^{10} 6p^1     $ &       258 297 & & $-$508 &   $-$182 &  $-$237 & &   17 & $-$67 \\
\hline
\end{tabular}
\label{tab:Tl-GRPP}
\end{table}

\begin{table}[h]
\caption{SCF excitation energies for Lu. See the caption of Table~\ref{tab:Ra-GRPP}.}
\begin{tabular}{lrrrrrrrr}
\hline
& \multicolumn{1}{c}{\textbf{Excitation energy}} & & \multicolumn{3}{c}{\textbf{Contributions from}} & & \multicolumn{2}{c}{\textbf{Errors of}} \\
      \cmidrule{2-2} \cmidrule{4-6} \cmidrule{8-9}
$\ldots 4f^{14} 6s^2 5d^1\; \rightarrow$ 
                   & \textbf{DCB+QED, finite nucl.}    & & \textbf{QED} & \textbf{Finite nucl.} & \textbf{Breit} & & \textbf{full GRPP}  & \textbf{valence GRPP} \\

 \hline
 $\ldots 4f^{14} 6s^2 6p^1     $ &    2463 & &   59 &      9 &   109 & &  2 &     1 \\
 $\ldots 4f^{14} 6s^1 5d^1 6p^1$ &   18388 & &  $-$64 &    $-$13 &    $-$9 & &  1 &   $-$10 \\
 $\ldots 4f^{14} 6s^1 5d^2     $ &   21217 & &  $-$92 &    $-$16 &   $-$83 & & $-$2 &   $-$14 \\
 $\ldots 4f^{14} 6s^1 6p^2     $ &   28915 & &  $-$18 &     $-$7 &   108 & &  4 &   $-$13 \\
 $\ldots 4f^{14} 6s^2          $ &   35136 & &   83 &     14 &    82 & &  4 &    $-$4 \\
 $\ldots 4f^{14} 6s^1 5d^1     $ &   46334 & &  $-$55 &    $-$10 &   $-$35 & &  1 &   $-$15 \\
 $\ldots 4f^{14} 6s^1 6p^1     $ &   62535 & &   $-$7 &     $-$4 &    84 & &  4 &   $-$20 \\
 $\ldots 4f^{14} 5d^2          $ &   66638 & & $-$167 &    $-$30 &  $-$135 & & $-$1 &   $-$27 \\
 $\ldots 4f^{14} 6s^1          $ &  136744 & &    6 &     $-$1 &    31 & &  5 &   $-$38 \\
 $\ldots 4f^{14} 5d^1          $ &  144088 & & $-$148 &    $-$28 &   $-$95 & &  2 &   $-$48 \\
\hline
\end{tabular}
\label{tab:Lu-GRPP}
\end{table}

\begin{figure}[htbp]
\begin{center}
\includegraphics[width=0.5\columnwidth]{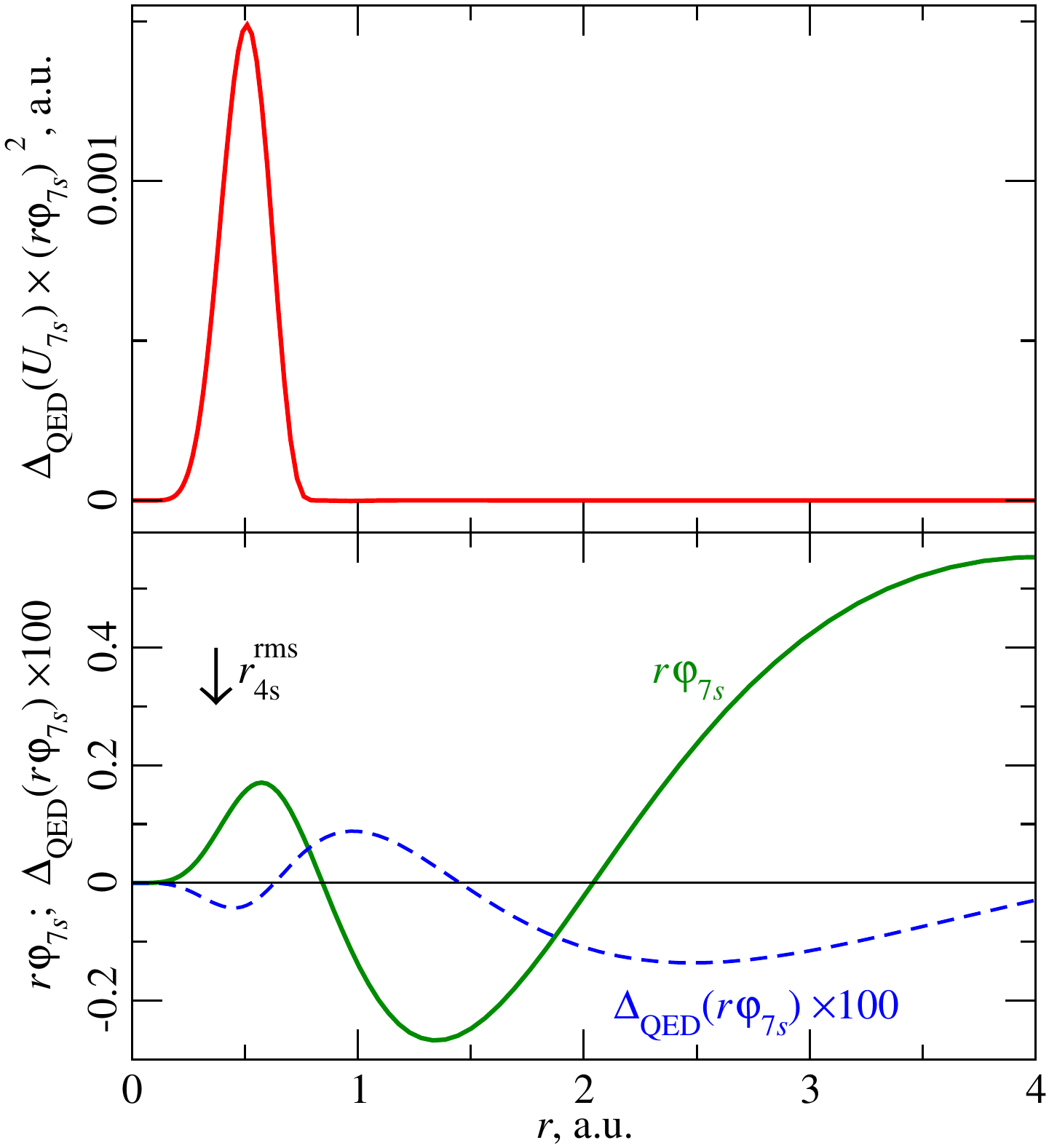}
\end{center}
\caption{Radial part of the $7s$ pseudospinor of Ra, $r\varphi_{7s}(r)$; QED effect on $r\varphi_{7s}(r)$, $\Delta_{\rm QED}(r\varphi_{7s})$; ``radial density'' of QED energy shift for the $7s$ pseudospinor, $\Delta_{\rm QED}(U_{7s})\times(r\varphi_{7s})^2$. The vertical arrow marks the root-mean-square radius of the outermost excluded (inner-core) $s$-spinor, $r^{\rm rms}_{4s}$.}
\label{fig:raqed}
\end{figure}

To be able to estimate the QED contributions to various results of calculations, we also performed the GRPP generation without introducing the QED model potential into the reference all-electron atomic SCF equations, thus obtaining the consistent QED-free counterparts of our GRPPs. Due to the ``hardness'' of the present GRPP form~\cite{Mosyagin:20}, QED contributions to atomic pseudospinors are localized essentially 
outside of the inner-core area (Fig.~\ref{fig:raqed}).  The differences between the $U_{nlj}(r)$ functions obtained with and without accounting for QED effects, $\Delta_{\rm QED}(U_{nlj})$, can be considered as partial effective QED potentials. These potentials are naturally much less localized in the vicinity of the atomic nuclei than their analogs used in all-electron calculations. It might be instructive to consider the localization of functions $\Delta_{\rm QED}(U_{nlj}(r))\times(r\varphi_{nlj}(r))^2$, where $r\varphi_{nlj}(r)$ denotes the radial part of the pseudospinor $nlj$. These functions which can be interpreted as the ``radial density'' of QED energy shift for atomic pseudospinors, are localized approximately at the same distances from the nuclear center as the innermost wave of the corresponding pseudospinor (or equivalently, as the innermost pseudospinor with the same values of $l$ and $j$), being relatively large in the area where the radial part cannot be small. An example is provided in Fig.~\ref{fig:raqed}. This feature seems essential for numerical stability of GRPP-based estimates for QED effects on electronic state energies.

\subsection{Intermediate-Hamiltonian Fock space coupled cluster method: incomplete main model spaces}
\label{sec:intham}

Consider first a single Hilbert space and choose the model subspace $\LP$ in such a way that the target eigenstates  $\{\Psi_\mu\}, \mu=1,\dots,M$ of the electronic Hamiltonian $H$ have large projections onto $\LP$,   and the dimension $D$ of $\LP$ is larger than the number of target states $M$. We shall denote the projectors onto $\LP$ and its orthogonal complement $\LQ$ by $P$ and $Q$, respectively. 

A state-selective effective Hamiltonian (intermediate Hamiltonian) $\HINT$ and the corresponding wave operator $\Omega$ are defined by the requirements~\cite{Malrieu:85,Durand:87} 
\begin{equation}
\HINT\MV_\mu=E_\mu\MV_\mu, \quad\quad \Omega \MV_\mu=\Psi_\mu, \quad\quad \mu=1,\dots, M;\;\;M<D,
\label{defini}
\end{equation}
where $\{E_\mu\}$ denote the eigenvalues of $H$ corresponding to the target eigenvector $\Psi_\mu$ and $\{\MV_\mu\}$ are the corresponding eigenvectors of  $P\HINT P$ (target model vectors). This means that the conventional Bloch equation must hold only within the subspace of $\LP$ spanned by the target model vectors. Denoting the orthogonal projector onto this subspace by $\PPM$, one can write this requirement as
\begin{equation}
(\Omega P \HINT- H\Omega )\PPM= 0
\label{projectedbloch}
\end{equation}
or in the form of shifted Bloch equation~\cite{Mukhopadhyay:92,Zaitsevskii:92},
\begin{equation}
(\Omega P \HINT- H\Omega + W)P= 0, 
\label{shiftedbloch}
\end{equation}
where  $W$ is a rather arbitrary operator meeting the condition  
\begin{equation}
W\PPM=0.
\label{bufferonly}
\end{equation}

Let us split the total Hamiltonian, normal-ordered with respect to the Fermi vacuum, into the one-electron part $H_0$ and the perturbation $V= H-H_0$ and suppose that the model space is spanned by a subset of the eigenfunctions of $H_0$. Provided that the intermediate normalization of $\Omega$ is assumed, $P\Omega P=P$, Eq.~(\ref{shiftedbloch}) is readily converted into
\begin{equation}
P\HINT P-PH_0P\equiv P\VINT P =PV\Omega P -PWP 
\label{hint}
\end{equation}
and
\begin{equation}
Q[\Omega,H_0]P +QWP=Q( 
V \Omega - 
\Omega P\, \VINT)P .
\label{waveoperator}
\end{equation}
A proper choice of $W$ should avoid numerical instabilities while solving Eqs~(\ref{hint}-\ref{waveoperator}) and prevent the appearance of large cluster amplitudes which could ruin the restricted-range cluster approximation for the wave operator.  
For instance, Refs.~\cite{Mukhopadhyay:92,Zaitsevskii:92} advocated the choice
\begin{equation}
W=Q[\Omega,\mathcal{S}]= Q\Omega \mathcal{S},\quad\quad
\mathcal{S}=P\mathcal{S}(P-\PPM),
\label{mm}
\end{equation}
where $\mathcal{S}$ is an energy-like shift operator. In this case, Eq.~(\ref{waveoperator}) can be presented in the form
\begin{equation}
Q[\Omega,(H_0+\mathcal{S})]P=QV\Omega P-Q\Omega PV\Omega P,
\label{mmbloch}
\end{equation}
i.\ e. (\ref{mm}) corresponds  to an uncompensated modification of $H_0$ within the model space. This modification can be straightforwardly used to suppress the effect of intruder states. 

Due to the requirement (\ref{bufferonly}), Eqs.~(\ref{hint}-\ref{waveoperator}) implicitly involve the unknown projector $\PPM$,
so that a computational scheme based directly on solving these equations would be rather cumbersome~\cite{Mukhopadhyay:92}. A more practical approach \cite{Landau:00,Landau:01,Eliav:05} implies the pre-partitioning of the model space into the main subspace $\LM$ projected by $P_M$ and intermediate subspace $\LI$ with its projector $P_I=P-P_M$,  both spanned by appropriate sets of Slater determinants, in such a way that the intermediate subspace contains all determinants with dangerously high zero-order energies and the contributions of intermediate-subspace determinants to all target eigenstates of $\HINT$ are rather small, i.~e.
\begin{equation}
\PPM P_I \approx 0.
\label{required}
\end{equation} 
The basic approximation of this approach consist in replacing the requirement (\ref{bufferonly}) by that involving the  projector onto the pre-defined main model space:
\begin{equation}
WP_M=0,\;\mbox{ or }\;W=WP_I .
\label{approxbufferonly}
\end{equation}
     
It is worth underlining that the model space separation satisfying the requirement (\ref{required}) differs essentially from that introduced by Malrieu et al. \cite{Malrieu:85,Durand:87}. In the latter case the main model space size should coincide with the number of target states, ${\rm tr}\;P_M= {\rm tr}\; \PPM$; in contrast, a reasonable accuracy of the approximation (\ref{required}) for $P_I$ projecting on a linear span of a set of Slater determinants normally implies that the main model space size is larger (and sometimes much larger) than the number of target states, ${\rm tr}\; P_M>{\rm tr}\; \PPM$.

In the present work we adopt the approximation~(\ref{required})--(\ref{approxbufferonly}). However, in contradistinction 
to the computational scheme developed and employed in Refs.~\cite{Landau:00,Landau:01,Eliav:05}, the present approach does not require 
the completeness of the main model space. In practice,
a rather small complete main model space can suffer from large deviations from the requirement (\ref{bufferonly}) whereas its extension rapidly increases its spreads in energy resulting in persistent intruder states. The use of incomplete $\LM$ largely eliminates this difficulty. In numerous cases, $\LM$ can be chosen simply as the linear span of a subset of model-space determinants with the lowest values of $H_0$; alternatively, restricted-active-space-like main model spaces are used in applications.  The total model space $\LP$ is still supposed to be complete and thus unambiguously defined by a set of ``active’’ spinors. The completeness of $\LP$ is essential for conserving the relative simplicity and transparency of the conventional FS-RCC computational scheme.

Let us turn to the Fock space coupled cluster scheme of constructing intermediate Hamiltonians and focus on the target sector $(m,n)$ of the Fock space ($m$ and $n$ denote the number of holes and particles, respectively). The wave operator is written in the normal-ordered exponential form,
\begin{equation}
\Omega=\left\{e^T\right\}, \quad\quad T=\!\!\sum_{\scriptsize\begin{array}{c}m’\le m \\ n’\le n\end{array}}T^{(m’,n’)}
\label{expt}
\end{equation}
where $T^{(m’,n’)}$ stands for the cluster operator component with $m’$ hole and $n’$ particle destruction operators. We shall follow the conventional scheme extracting the $T^{(m’,n’)}$ amplitudes with $m’<m$ or/and $n’<n$ from the cluster equations for the corresponding sectors. Suppose first that the coupled-cluster problem in the lower $(m’,n’)$ sectors are solved using the conventional Bloch equation, $W^{(m’,n’)}=0$. A rather general modification (\ref{shiftedbloch}) of the Bloch equation for the target sector can be defined in the following way:   
\begin{equation}
W^{(m,n)}=
\!\!\!\!\!\!\!\!
\sum_{
\begin{array}{c}
\scriptstyle L,\;K \vspace{-1ex}\\
\scriptstyle K:\, A_K^\dag|\Phi_0\rangle \in \LI^{(m,n)}   
\end{array}
} 
\!\!\!\!\!\!\!\!
 A^\dag_L \, t_{LK}^{(m,n)}S_{LK}A_K .
\label{ourw}
\end{equation}
Here $t_{LK}^{(m,n)}$ is the cluster amplitude and $S_{LK}$ is the shift value associated with the excitation $A^\dag_LA_K$ ($A_K$ and $A^\dag_L$ stand for chains of destruction and creation operators respectively); $\ket{\Phi_0}$ denotes the Fermi vacuum. We make use of the one-to-one correspondence between the set of quasiparticle destruction operator sets in $T^{(m,n)}$ and model determinants of the $(m,n)$ sector. Due to the restriction of summation over $K$, the action of $W^{(m,n)}$ is equivalent to that of $Q^{(m,n)}W^{(m,n)}P_I^{(m,n)}$.

Substituting (\ref{expt}) and (\ref{ourw}) into Eqs.~(\ref{hint}-\ref{waveoperator}) for the sector $(m,n)$, assuming the validity of the basic approximation~(\ref{required}) and applying the Wick theorem, one notices that all explicitly disconnected terms arising from lower-rank amplitudes cancel each other in a usual way~\cite{Lindgren:87} and arrives at the amplitude equations
\begin{equation}
Q^{(m,n)}\left([T^{(m,n)},H_0]P^{(m,n)}+W^{(m,n)}\right)P^{(m,n)}=Q^{(m,n)}(\contraction[0.5ex]{}{V}{}{\Omega}V\Omega-\contraction[0.5ex]{}{\Omega}{\,}{V}\Omega\, V_{\rm int} ) P^{(m,n)}
\label{tequation}
\end{equation}
or
\begin{equation}
t_{LK}^{(m,n)}=
\left\{
\begin{array}{l}
D_{LK}^{-1}\left(\contraction[0.5ex]{}{V}{}{\Omega}V\Omega-\contraction[0.5ex]{}{\Omega}{\,}{V}\Omega\, V_{\rm int} \right)_{LK}
,
\mbox{ if }A_K^\dag \vert \Phi_0\rangle\in \LM^{(m,n)}
\\ \\
(D_{LK}+S_{LK})^{-1}\left(\contraction[0.5ex]{}{V}{}{\Omega}V\Omega-\contraction[0.5ex]{}{\Omega}{\,}{V}\Omega\, V_{\rm int} \right)_{LK}
\mbox{ otherwise.}
\end{array}
\right.
\label{tlkequation}
\end{equation}
Here $D_{LK}$ is the conventional energy denominator associated with the excitation $A_LA^\dag_K$. The Wick contraction symbols denote only the formal connectivity, i.\ e. the terms marked by this symbol would be connected if $W^{(m,n)}$ and thus $T^{(m,n)}$ were connected. The choice of $W^{(m,n)}$ employed in the present work should destroy the connectivity (and thus the exact size consistency of results).

The choice of numerical parameters $S_{LK}$ playing the role of denominator shifts, on the one hand, should prevent the appearance of ill-determined energy denominators in (\ref{tlkequation}). On the other hand, it is desirable to reduce the difference in the treatment of excitations from main and low-energy intermediate determinants, thus reducing the errors due to inevitable deviations of (\ref{required}) from the exact equality. In the present work, we use ``imaginary’’ adjustable shifts (cf. Ref~\cite{Oleynichenko:20cpl})
%
\begin{eqnarray}
S_{LK}&=&i\,s_{K}\left(\frac{|s_K|}{|D_{LK}+i\,s_K|}\right)^{p}, \nonumber \\
s_K&=&
\overline{E}_{0M}
- \langle \Phi_0  \vert A_K H_0 A_K^\dag \vert\Phi_0\rangle ,
\label{choiceofs}
\end{eqnarray}
%
where $p$ is an integer non-negative parameter and $\overline{E}_{0M}$ denotes an energy value lying within the range of $P_M^{m,n}H_0P_M^{m,n}$ eigenvalues; normally we assumed that $\overline{E}_{0M}=\max_{K’\in \LM} \langle \Phi_0  \vert A_{K’} H_0 A_{K’}^\dag \vert\Phi_0\rangle$.
In this case the shift amplitude $s_K$ is simply the separation in ``unperturbed'' energy between the intermediate determinant $A^\dag_K\vert \Phi_0\rangle$ and the top of 
the main model space spectrum, so that for $p=0$ the real counterpart of $S_{LK}$, losing the dependence on $L$, would simply lift down its unperturbed energy of the intermediate-space determinant $A^\dag_K\vert\Phi_0\rangle$ to the highest $H_0$ value within the main model space. One readily realizes that $|S_{LK}|$ are relatively small for low-energy intermediate determinants and, if $p> 0$, for any high-energy excitations $K\to L$ corresponding to well-defined energy denominators. The rate of shift attenuation with the increase of $|D_{LK}|$ is determined by the $p$ value \cite{Zaitsevskii:17,Zaitsevskii:18a,Oleynichenko:20cpl}. 

It is to be underlined that the present form of $W^{(m,n)}$ is not separable and one could not hope to maintain the exact size consistency with truncated $T$ expansions even if the basic approximation (\ref{required}) was perfectly good. Therefore it seems advantageous to reduce shifts via using large attenuation parameters $p$ whenever it is possible without ruining the stability of 
the computational procedure or producing unacceptably large cluster amplitudes, aiming at avoiding large deviations from the conventional effective-Hamiltonian size-consistent scheme.

Finally, to avoid the risk of obtaining complex $\HINT$ eigenvalues, we replace the complex factor $(D_{LK}+S_{LK})^{-1}$ in Eq.~(\ref{tlkequation}) by its real part (\cite{Oleynichenko:20cpl}, cf.~\cite{Forsberg:97,Witek:02}).  

The intruder state problem for lower Fock space sectors is normally less severe than for higher ones; for instance, in many cases coupled cluster equations for the $(0h1p)$ can be solved using the Jacobi method without invoking the intermediate Hamiltonian machinery whereas for the $(0h2p)$ one this is quite rarely possible. Nevertheless, situations, where the shifting is required for some 
sectors with a smaller number of quasiparticles than that in the target sector, are not exotic. In such cases, additional terms would appear in Eq.~(\ref{tequation}) even if $W$ operators satisfy Eq.~(\ref{bufferonly}) exactly \cite{Mukhopadhyay:92}. Provided that $W$ operators for lower sectors are chosen so that
\begin{equation}
\forall \; m’\le m,\; n’\le n\quad W^{(m’,n’)}P_M^{(m,n)}=0
\label{consistent}
\end{equation}
these terms should not affect the main model states in the limit of exact solutions; following Refs.~\cite{Landau:00,Landau:01,Eliav:05}, we shall simply omit them. The condition (\ref{consistent}) is readily fulfilled for complete $\LM$ via using the universal partitioning of active spinor set into the main and intermediate subsets for all sectors, but it requires some check for general incomplete main subspaces. For the target $(0h2p)$ sector considered in the applications described below this check remains simple: an acceptable $W^{(0h1p)}$ cannot destruct a particle on a spinor occupied in any determinant from $\LM^{(0h2p)}$.

\section{Pilot applications}
\label{sec:results}

In order to further assess the accuracy of the new GRPP model and to get insight into the prospects and problems of its use in atomic and molecular excited state studies, we perform a series of pilot calculations on electronic transitions in two-valence atomic systems Ra, Tl$^+$, and Lu$^+$ using the intermediate-Hamiltonian FS-RCC scheme described above for correlation treatment. In all cases, the target states are considered as belonging to the $(0h2p)$ sectors with respect to closed-shell vacuum states defined by the Hartree--Fock configurations of these systems with two removed electrons. We focused on low-energy transitions involving the changes in $s$-subshell occupancies which are expected to be significantly affected by QED effects.
In spite of the formal similarity of ground-state configurations ($\dots s^2$), excited state nature and correlation patterns in
the three systems are radically different. The valence subsystem of Ra is well separated from the rather weakly polarizable $\dots 6s^26p^6$ closed-shell core. The correlations of the outermost subvalence $5d$ shell in Tl$^+$ (isoelectronic with the transition-metal Hg atom) with valence electrons  are expected to be much more important and complicated. The Lu$^+$ ion is characterized by a dense manifold of low-lying many-electron levels; furthermore, the closed $4f$ shell treated as subvalence in the present calculations is normally considered as valence one in lanthanide chemistry, so 
the results should be particularly sensitive to the level of treatment of correlations involving this subshell.

All pilot applications are made with the help of ``molecular'' software (the DIRAC19 code~\cite{DIRAC_code:19,DIRAC:20} for solving the SCF problem and the integral transformation 
step, as well as EXP-T~\cite{EXPT:22,Oleynichenko:EXPT:20} program at the FS-RCC stage) and employed conventional Gaussian bases, so that analogous studies can be readily performed for small molecules. Owing to the current software limitations, only semilocal valence GRPP components rather than the full nonlocal GRPPs are employed; for the transitions considered, the losses in accuracy due to this replacement are expected to be much smaller than 
the QED contributions to transition energies.
A more serious restriction arises from the inability of the present DIRAC versions to evaluate RPP integrals with high angular momentum ($l\ge{}7$) functions ($k,\;l,$ etc.); in some cases we tried to roughly estimate the errors due to the lack of such basis functions using all-electron models.

Since full FS-RCCSDT calculations remain feasible only with very restricted one-electron bases, we invoked the additive scheme and
estimated the contribution of triple excitations as the differences of FS-RCCSDT and FS-RCCSD transition energies using contracted versions of Gaussian bases. These bases were composed mainly of scalar-relativistic averaged atomic natural orbitals (ANOs); density matrices to be averaged were obtained in the series of single-reference CCSD or CCSD(T) calculations using the CFOUR software~\cite{CFOUR}. 

\subsection{Radium atom}
\label{sec:radium}

\begin{figure}
    \centering
    \includegraphics[width=0.6\columnwidth]{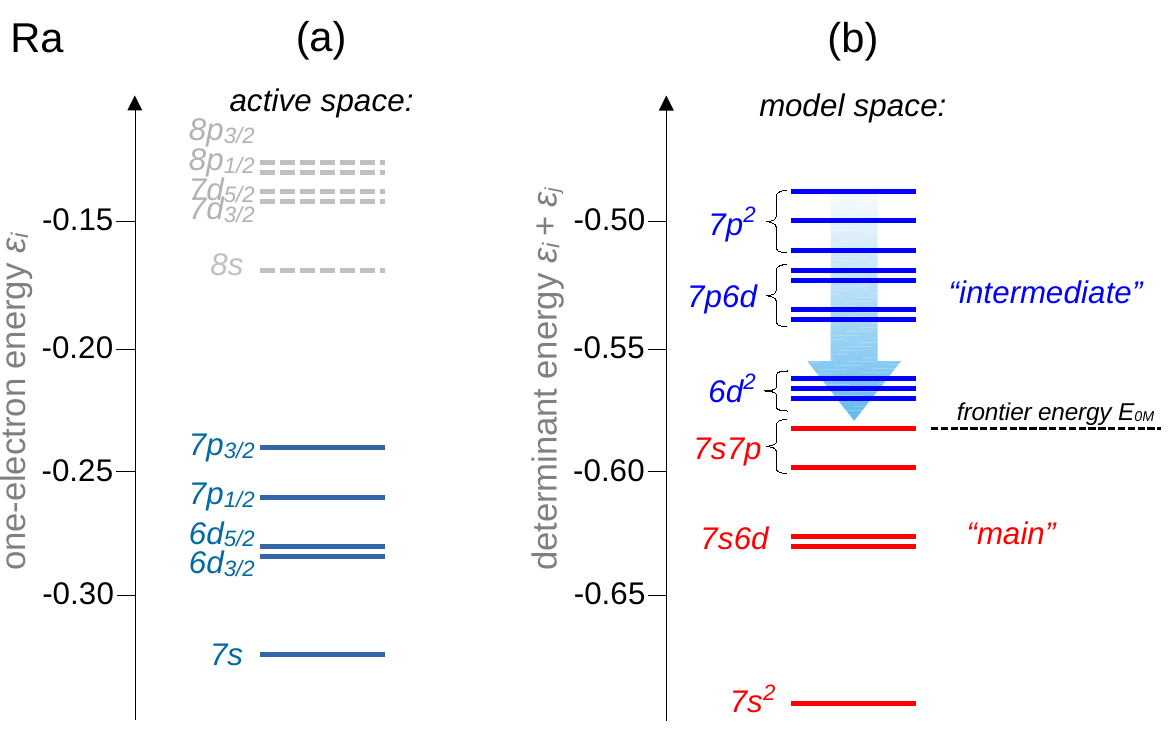}
    \caption{Model space for the Ra atom FS-RCC calculation: (a) energy diagram for one-electron spinors of the Ra$^{2+}$ ion; (b) energy diagram for model space determinants in the $(0h2p)$ sector (``unperturbed'' spectrum of the $H_0$ operator). Target electronic states must be dominated by ``red'' determinants. The energy denominator shifting (Eq.~(\ref{choiceofs})) is performed for excitations corresponding to intermediate space determinants (blue), while amplitudes of excitations from the main space (red) are calculated with original (unshifted) denominators (see Eq.~\ref{tequation}).}
    \label{fig:radium-model-space}
\end{figure}

The present calculations aimed to model the low-lying states  $7s^2\;^1S$, $7s6d \;^3D_{1,2,3}$, $7s6d \,^1D_{2}$, $7s7p \;^3P^o_{0,1,2}$ and $ 7s7p \;^1P^o_1$ of the neutral radium atom. An active spinor set defining the total model space should include at least $7s$, $6d$, and $7p$ spinors and the main model space was naturally chosen as the linear span of $7s^2$, $7s6d_{3/2,5/2}$ and $7s7p_{1/2,3/2}$ determinants whereas $7p^2$, $7p6d$ and  $6d^2$ states were treated as intermediate (active and model spaces together with the shifting scheme are schematically depicted on Fig.~\ref{fig:radium-model-space}). The shift parameters $s_{K}$ were chosen according to Eq.~(\ref{choiceofs}), assuming that $\overline{E}_{0M}$ coincides with the top of main model space part of $H_0$ spectrum, i.\ e. with the ``unperturbed'' energy of the $7s7p_{3/2}$ determinants; the attenuation parameter value was usually $p=3$. The basis set used to expand the components of pseudospinors comprised the $s$, $p$, and $d$ subsets of uncontracted Dyall’s  quadruple-zeta set \cite{Dyall:09} adapted to the GRPP model, additional diffuse functions and even-tempered sequences of higher angular momentum functions (cf.~Ref~\cite{Landau:00}). The overall basis set size was ($17s14p12d11f8g7h6i)$; exponential parameters are provided in Supplementary materials.

The accuracy of intermediate-Hamiltonian FS-RCCSD energy estimates obtained with minimum model space, hereafter denoted as ($6d7sp$), leaves much to be desired (Table~\ref{tab:ra}). The deviations from experimental excitation energies are several times greater than the corresponding QED corrections, being comparable to those of effective-Hamiltonian incomplete-model-space FS-RCCSD calculations with the Dirac-Coulomb-Breit Hamiltonian (without QED corrections) \cite{Landau:00}. It is logical to suppose that the use of minimum model space implies the necessity to go beyond the CCSD approximation. Unfortunately, the attempt to estimate the contributions of non-perturbative triples from FS-RCCSDT calculations in smaller basis sets of atomic natural orbital failed. In contrast with the corresponding FS-RCCSD problem, no convergent solutions for the FS-RCCSDT 
were obtained with the chosen $s_{K}$ and $p$ parameters; increasing $|s_{K}|$ or decreasing $p$ moderately, one arrives at the solutions which should be interpreted as incompatible with the cluster hypothesis (huge amplitudes of double excitations acting only on the buffer-subspace determinants $6d^2$, never appeared in FS-RCCSD calculations). 

The valence and subvalence shells in neutral Ra are rather well separated both spatially and energetically, so that no intruder state with subvalence hole(s) is expected to appear. In this situation the simplest and rather economical way to improve the accuracy might consist in enlarging the $(0h2p)$ ``buffer'' subspace~\cite{Landau:00,Landau:01}. We used the extended model space arising from the active spinor set $\{5f$, $6d$, $7spd$, $8sp\}$, while the main model space was defined as previously. 
Similar to the case of minimum model space, no convergence problem was encountered in the embedded $(0h1p)$ sector; however, to prevent the appearance of rather large amplitudes in $T^{(0h1p)}$, the shifts were also applied in this sector, assuming that the model determinants with occupied $7s$, $7p_{1/2,3/2}$, and $ 6d_{3/2,5/2}$ spinors span the main subspace (and thus fulfilling the requirement (\ref{consistent})) and using the conventional rule of intermediate-state shifting (Eq.~(\ref{choiceofs}). The $\overline{E}_{0M}$ value was chosen as the $H_0$ eigenvalue for the uppermost ``main''  determinant $[\dots]7s7p_{3/2}$). The extension of the model space improves the results radically; the deviations of FS-RCCSD excitation energies from their experimental counterparts became less than a hundred wavenumbers. 
To estimate the contribution of triple excitations defined with respect to the new model space, we performed FS-RCCSDT / FS-RCCSD calculations using a necessarily limited basis of scalar-relativistic ANOs $[6s6p6d4f3g2h]$ augmented with the difference of spatial parts of Hartree-Fock $5p_{1/2}$ and $5p_{3/2}$ spinors. Only 20 electrons were correlated ($5s5p$ subshells frozen after the Hartree--Fock stage) and virtual spinors with energies above 8 a.u. were rejected. At first sight, the FS-RCCSD transition energies corrected for triples (SD+$\Delta_{\rm T}$, see Table~\ref{tab:ra}) are less accurate than the original FS-RCCSD values. However, the agreement of the latter values with the experimental data appears partially fortuitous, whereas the bulk of errors in FS-RCCSD+$\Delta_{\rm T}$ estimates have a clear origin in the lack of high-angular-momentum functions in the employed Gaussian basis and in removing the core shells up to $4spdf$ inclusively within the current GRPP model. 
A series of all-electron calculations demonstrated that the addition of three sets of $k$ functions lowered the FS-RCCSD $7s^2-6d7s$ excitation energy estimates by $20\div25$ cm$^{-1}$, leaving nearly unchanged the $7s^2-7s7p$ ones. The correlations involving the $n=4$ shells push the calculated $7s^2-6d7s$ and $7s^2-7s7p$ excitation energies up by $25\div58$ cm$^{-1}$ and $75\div85$ cm$^{-1}$ respectively. 

The QED contributions to excitation energies evaluated as differences between the values obtained 
using the GRPP generated with and without accounting for the QED effects
agree quantitatively with their counterparts resulting from high-level all-electron calculations of Ginges et al.~\cite{Ginges:15}.

\begin{table}[htp]
\caption{Experimental excitation energies (EE)~\cite{Sansonetti:05}, deviations of the calculated EE from their experimental counterparts
 and contributions to EE from QED effects ($\Delta_{\rm QED}$) for the Ra atom, cm$^{-1}$.  
}
\begin{center}
\begin{tabular}{llrrrrrrrrrr}
\hline
$7s^2\; ^1S_0\to$& \multicolumn{1}{c}{Exptl.} 
                          &  \multicolumn{1}{c}{SD$^{a)}$}     
                                     &  \multicolumn{1}{c}{SD$^{a)}$}   
                                            &\multicolumn{1}{c}{SD$+\Delta_{\rm T}^{a)}$}
                                                       & \multicolumn{3}{c}{Other}                & & \multicolumn{2}{c}{$\Delta_{\rm QED}$ }   \\
\cmidrule{6-8}\cmidrule{10-11}                                                                                    
                 &        & ($6d7sp$) &\multicolumn{2}{c}{($5f6d7spd8sp$)}   &  RCC$^{b)}$ & FS-RCC$^{c)}$ & LCC-CI$^{d)}$& & SD$^{a,e)}$  &  LCC-CI$^{d)}$\\  \hline 
$7s7p \;^3P^o_0$ & 13 078 &  102    & \ 9   & 51       &   75   &105      & \ \ 58   & & $\ -66$ & $-$64 \\
$7s6d \;^3D_1$   & 13 716 &  $-$203 & 10    & $-$110   & $-$204 &111      &  $-$63   & & $-107$  & $-$92 \\
$7s6d \;^3D_2$   & 13 994 &  $-$185 & 14    & $-$109   & $-$181 &108      & \ \ 31   & & $-101$  & $-$91 \\
$7s7p \;^3P^o_1$ & 13 999 &  110    & 12    &    45    &   86   &103      & \ \ 41   & & $\ -66$ & $-$64 \\
$7s6d \;^3D_3$   & 14 707 &  $-$126 & 43    & $-$75    & $-$151 &132      & \ \ 29   & & $\ -75$ & $-$85 \\
$7s7p \;^3P^o_2$ & 16 689 &  194    & 36    &  44      &   169  &116      & \ \ 73   & & $\ -61$ & $-$60 \\
$7s6d \;^1D_2$   & 17 081 &  55     & 99    & $-$17    &   566  &\ 78     &\ \ 181   & & $\ -97$ & $-$97 \\
$7s7p \;^1P^o_1$ & 20 715 &  324    & 66    &  9       &   426  & 107     & $-$110   & & $\ -60$ & $-$61 \\
\hline
\end{tabular}
\end{center}
\raggedright
\footnotesize
$^{a)}$ present work, FS-RCCSD (SD) and FS-RCCSD corrected for triple contributions within the additive scheme (SD$+\Delta_{\rm T}$). The FS-RCC active space composition is indicated in parentheses,\\
$^{b)}$ incomplete model space multireference effective-Hamiltonian RCCSD
and \ 
$^{c)}$ intermediate-Hamiltonian FS-RCCSD all-electron calculations with Dirac--Coulomb--Breit Hamiltonian~\cite{Landau:00};\\
$^{d)}$ all-electron calculations using the combination of correlation potential, linearized singles-doubles coupled-cluster, and the configuration interaction methods, Ref.~\cite{Ginges:15};\\
$^{e)}$ active space ($5f6d7spd8sp$).
\label{tab:ra}
\end{table}

\subsection{Thallium cation Tl$^+$}
\label{sec:thallium}

We restricted our attention to four low-energy excitations ($6s^2\; ^1S_0 \to$ $6s6p\; ^3P^o_{0,1,2}\, , \; ^1P^o_1$ ) of Tl$^+$, so that the minimum model space was defined by eight active spinors ($6s$ and $6p_{1/2,3/2}$) and its main subspace was spanned by the determinants $6s^2$ and $6s6p_{1/2,3/2}$. The employed $(14s14p13d13f8g7h5i)$ primitive Gaussian basis (see Supplementary materials) was built essentially in the same way as that used for Ra. All explicitly treated electrons, including those of the $4spdf$ shell, were correlated in FS-RCCSD calculations. It is worth noting that the effect of correlations involving the shell with the main quantum number $n=4$ on Tl$^+$ transition energies reaches 540 cm$^{-1}$, so that the conventional small-core RPP models with 60 excluded core electrons \cite{Metz:00,Mosyagin:97} are hardly suitable for quantitative excited state modeling. The intermediate-Hamiltonian FS-RCCSD calculations with the minimum model space and shift parameters defined similarly to the previous case yielded rather poor excitation energy estimates (Table~\ref{tab:tl}), with 
a maximum error of about 400 cm$^{-1}$. The corrections accounting for triples in the cluster expansions were derived from the results of FS-RCCSD and FS-RCCSDT calculations within the basis including [6s7p6d5f4g3h2i] scalar-relativistic ANOs and the difference functions $4p_{1/2}-4p_{3/2}$, $4d_{3/2}-4d_{5/2}$, and $4f_{5/2}-4f_{7/2}$; only 20 (subvalence 
$5spd$ and valence) electrons were correlated. In contrast with the previous case, the switch from RCCSD to RCCSDT did not give rise to any convergence difficulties; however, the maximum absolute value of $T^{(0h2p)}_2$ amplitude increased significantly (by 30 \%), still remaining moderate (ca.~0.17). It should be noticed that, according to the results of similar calculations in somewhat smaller ANO bases, the correction $\Delta_{\rm T}$ for the open-shell singlet-like state $^1P^o_1$ is especially sensitive to the quality of high-angular momentum basis components and our best estimate was quite far from saturation. The incorporation of $\Delta_{\rm T}$ brought the calculated transition energies closer to their experimental counterparts; the remainder deviations, as can be seen in Table~\ref{tab:tl}, were smaller than the estimates of the QED contributions with the help of the GRPP ($\Delta_{\rm QED}$). The bulk of these deviations can be interpreted as arising from the underestimation of correlation energy lowering for singlet-like states with respect to that for triplet-like states, which is a rather usual consequence of the lack of high-angular-momentum basis functions. 
The comparison of FS-RCC and SCF results indicated a non-negligible effect of correlations on $\Delta_{\rm QED}$.

In contrast with the case of atomic Ra, one could not hope 
that a simple extension of the model space in the $(0h2p)$ Fock space sector should radically improve the results since the couplings to low-lying $(1h3p)$ determinants with $5d$ holes are expected to be at least as important as those to $(0h2p)$ ones. Indeed, it can be seen from Table~\ref{tab:tl} that the model space extension via adding the spinors $7s$,$7p$, and $6d$ to the active set did not bring any 
improvement to the resulting FS-RCCSD transition energies.

\begin{table}[htp]
\caption{Experimental excitation energies (EE)~\cite{Sansonetti:05}, deviations of the calculated EE from their experimental counterparts and contributions to EE from QED effects ($\Delta_{\rm QED}$)for the Tl$^+$ ion, cm$^{-1}$.
 }
\begin{center}
\begin{tabular}{llrrrrrrrr}
\hline
$6s^2\; ^1S_0\to$ & \multicolumn{1}{c}{Exptl.} &  \multicolumn{1}{c}{SD$^{a)}$}     &  \multicolumn{1}{c}{SD$^{a)}$} & \multicolumn{1}{c}{SD$+\Delta_{\rm T}^{a)}$}&\multicolumn{2}{c}{Other} & & \multicolumn{2}{c}{$\Delta_{\rm QED}$ } \\ 
\cmidrule{6-7} \cmidrule{9-10}
 &        & \multicolumn{1}{c}{($6sp$)} & \multicolumn{1}{c}{($6spd7sp$)}   &   \multicolumn{1}{c}{($6sp$)}        & \multicolumn{1}{c}{FS-RCC$^{b)}$} & \multicolumn{1}{c}{CI+MBPT$^{c)}$} & & \multicolumn{1}{c}{SD$^{a)}$}   &  \multicolumn{1}{c}{SCF$^{d)}$}     \\
\hline
$6s6p$ $^3P^o_0$ & 49 451 & $-$394  &  $-$433       &     $-$128         & 1212 / 17   & 867         & &$-$266     & $-$262  \\
$6s6p$ $^3P^o_1$ & 52 394 & $-$312  &  $-$363       &     $-$38          & 1378 / 70   & 714         & &$-$236     & $-$256  \\
$6s6p$ $^3P^o_2$ & 61 727 & $-$181  &  $-$295       &         8          & 1556 / 204  &1066         & &$-$215     & $-$259  \\
$6s6p$ $^1P^o_1$ & 75 663 &    272 &     173       &     127            & 2010 / 729  &$-1010$      & &$-$313     & $-$237 \\
\hline
\end{tabular}
\end{center}
\label{tab:tl}
\raggedright
\footnotesize
$^{a)}$ present work. See the footnotes in Table \ref{tab:ra} for notations; \\
$^{b)}$ all-electron FS-RCC calculations with Dirac--Coulomb--Breit Hamiltonian; two values correspond to two different vacuum state choices, Tl$^{+}$~$6s^2$ or Tl$^{3+}$~$6s^0$)~\cite{Eliav:96tl}; \\
$^{c)}$ combined configuration interaction / many-body perturbation theory correlation treatment,~\cite{Dzuba:96}; \\
$^{d)}$ present work, numerical SCF calculations using Dirac--Coulomb--Breit Hamiltonian with/without model QED operator added.
\end{table}

\subsection{Lutetium cation Lu$^+$}
\label{sec:lutetium}

The Lu$^+$ ion is a more complicated system than those considered above. The difficulty of modeling of the Lu$^+$ electronic states is mainly determined by the following circumstances. Firstly, the electronic states lying below 60000 cm$^{-1}$ comprised two electrons over the Lu$^{3+}$ closed-shell vacuum (the $(0h2p)$ Fock space sector) distributed over the $6s$, $6p$ and $5d$ spinors which are quite close in energy to each other; the resulting spectrum is relatively dense and contains dozens of electronic states. These states can be classified by the leading configuration, i.~e. as the $6s^2$, $6s5d$, $6s6p$, $5d^2$ and $6p5d$ states. Obviously, these configurations can be chosen as the target ones during the intermediate Hamiltonian FS-RCC calculation; the intruder states dominated by the $6p^2$ configuration are to be suppressed by appropriate denominator shifts (see Sect.~\ref{sec:intham}). Thus shifting amplitudes $s_K$ (see Eq.~\ref{choiceofs}) were calculated with respect to the energy $\overline{E}_{0M}$ of the $6p_{3/2}5d_{5/2}$ determinants. Note that care has to be taken to ensure that the ``amount'' of the main determinants is large enough (say more than 95\%) in all the target states. Secondly, the subvalence $5s$, $5p$ and the valence $4f$ electrons contribute significantly to the dynamic correlation and must be treated explicitly even when evaluating the corrections for triple amplitudes. The QED effects on excitation energies are expected to be of order $\sim 100\div250$ cm$^{-1}$. The order of contributions of the triple excitation amplitudes is expected to be similar to the above. On the contrary, the error introduced by the restriction to only the semilocal (valence) part of GRPP is not expected to exceed 30 cm$^{-1}$ for the states considered (see Table~\ref{tab:Lu-GRPP}), thus justifying the use of the semilocal part of RPP.

The basis set for the Lu atom was designed to be used in conjunction with the semilocal RPPs (28 electrons in core). Exponential parameters for the $s$, $p$ and $d$ functions were borrowed from the Dyall's quadruple-zeta basis set~\cite{Gomes:10}. Exponents roughly corresponding to the area of localization of the subvalence $(4$-$5)s$-, $(4$-$5)p$-, $5d$-shells, as well as the exponents describing the $4f$-shell, were re-optimized. Correlation $g$-, $h$- and $i$-functions were taken from the universal series of Malli et al \cite{Malli:93}. The final composition of the basis set was $(17s15p14d16f10g6h6i)$. FS-RCCSD excitation energies of the Lu$^{2+}$ cation (the $6s \rightarrow 6p$, $5d$, $7s$, $7p$, $6d$, $5f$ excitations were considered) and the neutral Lu atom (the $6s^25d$ $\rightarrow$ $6s^26p$, $6s^27s$, $6s^27p$, $6s^26d$, $6s^25f$ excitations) were found to be stable with respect to any further extension of the basis set. In order to perform calculations accounting for triples, the contracted ANO-type version $[7s9p7d6f4g3h2i]$ of the basis was constructed; density matrices to be averaged were obtained by the scalar relativistic CCSD(T) method. Contracted basis set was augmented by the differences of Hartree--Fock atomic spinors $4p_{3/2}-4p_{1/2}$, $5p_{3/2}-5p_{1/2}$ and $4d_{5/2}-4d_{3/2}$. All explicitly treated electrons were correlated in FS-RCCSD calculation performed with the large uncontracted basis set, while during the triple correction calculation step the $4s4p4d$ shells were kept frozen.

\begin{figure}
    \centering
    \includegraphics[width=0.5\columnwidth]{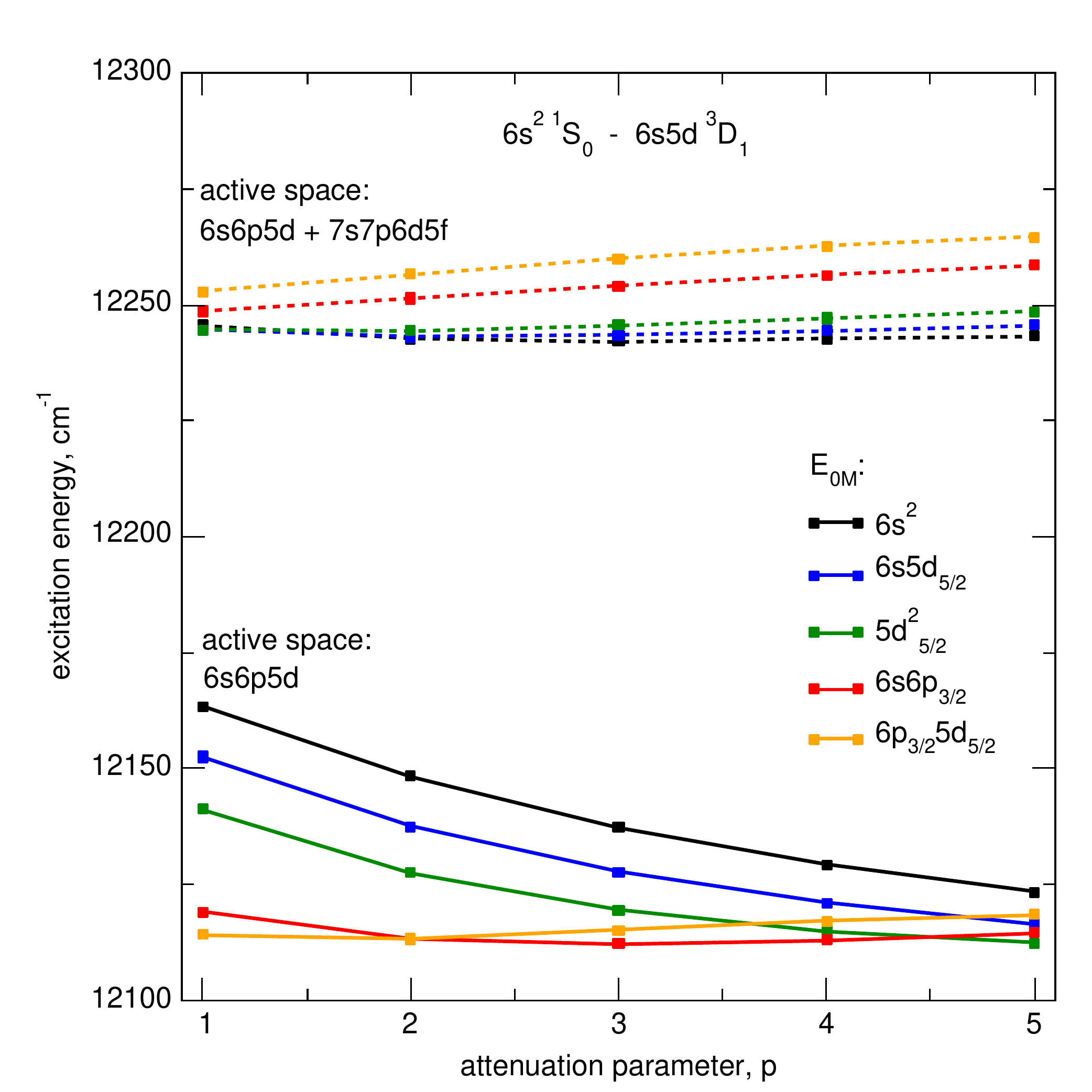}
    \caption{FS-RCCSD $6s^2\ ^1S_0 \rightarrow 6s5d\ ^3D_1$ transition energies in Lu$^+$ as functions of the attenuation parameter $p$ (see formula~(\ref{choiceofs})) (QED corrections are included; experimental value 11796 cm$^{-1}$). Calculations were performed with different choices of the active space (solid lines for $6s6p5d$ and dashed lines for $6s6p5d7s7p6d5f$) and ``frontier'' determinants defining $\overline{E}_{0M}$.
    }
    \label{fig:Lu-intham-3D1}
\end{figure}

\begin{figure}
    \centering
    \includegraphics[width=0.5\columnwidth]{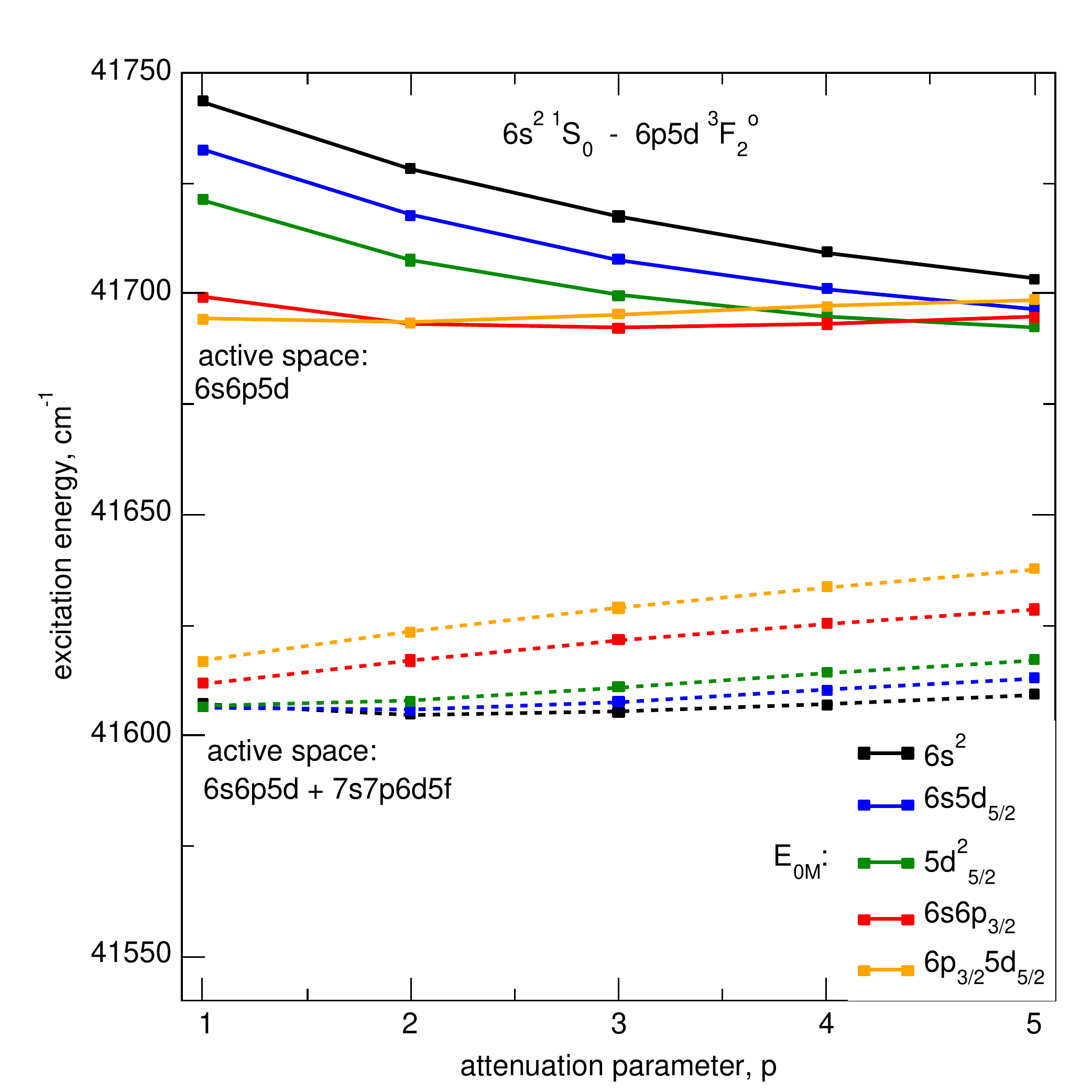}
    \caption{FS-RCCSD $6s^2\ ^1S_0 \rightarrow 6p5d\ ^3F_2^o$ transition energies in Lu$^+$  (experimental value 41225 cm$^{-1}$) as functions of the attenuation parameter $p$. See the caption for Figure~\ref{fig:Lu-intham-3D1} for explications.}
    \label{fig:Lu-intham-3F2}
\end{figure}

Since the lutetium cation possesses the most complicated electronic structure
among all systems considered in the work, we use this example to illustrate the stability of the new intermediate-Hamiltonian technique reported in Section~\ref{sec:intham} with respect to its parameters, the frontier energy $\overline{E}_{0M}$ and the attenuation parameter $p$. Figures~\ref{fig:Lu-intham-3D1} and~\ref{fig:Lu-intham-3F2} present the dependencies of calculated $6s^2\ ^1S_0 \rightarrow 6s5d\ ^3D_1$ and $6s^2\ ^1S_0 \rightarrow 6p5d\ ^3F_2^o$ transition energies in Lu$^+$ on these parameters.

One can clearly see that the results are pretty insensitive to the attenuation parameter $p$, provided the assumption that $\overline{E}_{0M}$ is the ``unperturbed'' energy of the uppermost model space determinants, namely, those corresponding to the $6s6p_{3/2}$ and $6p_{3/2}5d_{5/2}$ configurations. The variations of excitation energies with respect to $p$ do not exceed 10 cm$^{-1}$ and can be regarded as negligible. When extending active space by the next group of virtual shells, calculated excitation energy becomes nearly independent on $p$ for all choices of the frontier energy.
Quite similar patterns occur for Ra and Tl$^+$. This gives a hope that FS-RCC augmented with the intermediate Hamiltonian formulation proposed here can be used as a ``black-box'' method, which would require only setting the determinant composition of the main model space (or even the uppermost ``main'' determinant only). Obviously, this does not eliminate the need to check \emph{a posteriori} that target intermediate Hamiltonian eigenstates have sufficiently large $\mathcal{L}_I$ projections and that all cluster amplitudes are moderate. The applicability to molecular electronic states is yet to be studied in future research.

The results of our calculations are summarized in Table~\ref{tab:lu}. We also predict the existence of the two additional $5d^2$ states of Lu$^{+}$ which were not previously observed in spectroscopic experiments, namely, $^1G$ (the best theoretical value including all corrections for the corresponding energy level is 39229 cm$^{-1}$) and $^1S$ (47254 cm$^{-1}$).

\begin{table}[htp]
\caption{
Experimental excitation energies (EE)~\cite{Sansonetti:05}, deviations of the calculated EE from their experimental counterparts, and contributions to EE from QED effects ($\Delta_{\rm QED}$) for the Lu$^+$ ion, cm$^{-1}$.
}
\begin{center}
\begingroup \setlength{\tabcolsep}{4pt}
\begin{tabular}{cccrrrrrrrrrr}
\hline
\multicolumn{2}{l}{$6s^2\; ^1S_0\to$} & Exptl. & \multicolumn{1}{c}{SD$^{a)}$}     &  \multicolumn{1}{c}{SD$^{a)}$}           &\multicolumn{1}{c}{SD$+\Delta_{\rm T}^{a)}$}&\multicolumn{4}{c}{Other} & & \multicolumn{2}{c}{$\Delta_{\rm QED}$ } \\ 
 \cmidrule{7-10} \cmidrule{12-13}
& & & AS-1 & AS-2 & AS-1 & FS-RCC$^{b)}$ & CI+PT$^{c)}$ & MRCI$^{d)}$ & CI+SD$^{e)}$& & \hspace{-2ex}SD$^{a)}$ & CI+PT$^{c)}$     \\
\hline
$6s5d$ & $^3D_1$ & 11 796 & 322 & 468 & 358 & 558 & $-$132 & 245 & 152
 & & $-$139 & $-$144 \\
       & $^3D_2$ & 12 435 & 318 & 459 & 360 & 550 & $-$55 & 75 & 260 & & $-$138 & $-$143 \\
       & $^3D_3$ & 14 199 & 279 & 418 & 340 & 503 & 68 & $-$385 & 274 & & $-$130 & $-$134 \\
$6s5d$ & $^1D_2$ & 17 333 & 415 & 491 & 421 & 559 & 542 & $-$842 & 559 & & $-$151 & $-$160 \\
$6s6p$ & $^3P^o_0$ & 27 264 & $-$42 & $-$189 & $-$91 & $-$173 & 39 & 1400 & 393 & & $-$104 & $-$105 \\
       & $^3P^o_1$ & 28 503 & $-$38 & $-$172 & $-$82 & $-$63 & 17 & 1343 & 388
 & & $-$105 & $-$106 \\
       & $^3P^o_2$ & 32 453 & $-$16 & $-$174 & $-$78 & $-$159 & 150 & 1410 & 465 & & $-$99 & $-$97 \\
$5d^2$ & $^3F_2$ & 29 407 & 457 & 783 & 675 & --- & --- & --- & 344 & & $-$269 & --- \\
       & $^3F_3$ & 30 889 & 435 & 743 & 660 & --- & --- & --- & 349 & & $-$263 & --- \\
       & $^3F_4$ & 32 504 & 428 & 720 & 652 & --- & --- & --- & 481 & & $-$258 & --- \\
$5d^2$ & $^3P_0$ & 35 652 & 453 & 772 & 605 & --- & --- & --- & 21 & & $-$264 & --- \\
       & $^3P_1$ & 36 557 & 451 & 759 & 621 & --- & --- & --- & 17 & & $-$261 & --- \\
       & $^3P_2$ & 38 575 & 667 & 826 & 679 & --- & --- & --- & 627 & & $-$252 & --- \\
$5d^2$ & $^3P_2$ & 36 098 & 637 & 863 & 685 & --- & --- & --- & 465 & & $-$260 & --- \\
$6s6p$ & $^1P^o_1$ & 38 223 & 267 & 189 & 53 & 241 & $-$838 & 210 & --- & & $-$130 & $-$129 \\
$5d6p$ & $^3F^o_2$ & 41 225 & 474 & 413 & 269 & --- & --- & --- & 564 & & $-$243 & --- \\
       & $^3F^o_3$ & 44 919 & 499 & 412 & 258 & --- & --- & --- & 656 & & $-$235 & --- \\
       & $^3F^o_4$ & 48 537 & 535 & 402 & 248 & --- & --- & --- & 767 & & $-$228 & --- \\
$5d6p$ & $^1D^o_2$ & 45 459 & 602 & 443 & 283 & --- & --- & --- & 690 & & $-$234 & --- \\
$5d6p$ & $^3D^o_1$ & 45 532 & 1029 & 625 & 371 & --- & --- & --- & 483 & & $-$236 & --- \\
       & $^3D^o_2$ & 46 904 & 1003 & 603 & 368 & --- & --- & --- & 570 & & $-$239 & --- \\
       & $^3D^o_3$ & 48 733 & 1063 & 595 & 382 & --- & --- & --- & 626 & & $-$236 & --- \\
$5d6p$ & $^3P^o_0$ & 49 964 & 1126 & 635 & 344 & --- & --- & --- & 541 & & $-$236 & --- \\
       & $^3P^o_1$ & 50 049 & 1120 & 635 & 345 & --- & --- & --- & 567 & & $-$233 & --- \\
       & $^3P^o_2$ & 51 202 & 989 & 572 & 320 & --- & --- & --- & 682 & & $-$228 & --- \\
$5d6p$ & $^1F^o_3$ & 53 079 & 1601 & 632 & 462 & --- & --- & --- & --- & & $-$234 & --- \\
$5d6p$ & $^1P^o_1$ & 59 122 & 2528 & 737 & 529 & --- & --- & --- & --- & & $-$194 & --- \\
\hline
\end{tabular}
\endgroup
\end{center}
\label{tab:lu}
\raggedright
\footnotesize
$^{a)}$ present work. The FS-RCC active spaces are denoted as AS-1 ($5d6sp$) and AS-2 ($5df6spd7sp$). See also the footnotes in Table \ref{tab:ra}; \\
$^{b)}$ all-electron extrapolated intermediate Hamiltonian (XIH~\cite{Eliav:05}) FS-RCCSD calculations with DCB Hamiltonian and the $1s2s2p$ electrons kept frozen~\cite{Kahl:19}, QED was included using the model Lamb shift potential~\cite{Shabaev:13,Shabaev:18}; \\
$^{c)}$ combined configuration interaction / many-body perturbation theory calculations with DCB Hamiltonian using numerical SCF spinors~\cite{Kahl:19}, QED is included via the model potential of Flambaum et al.~\cite{Flambaum:05}; \\
$^{d)}$ all-electron GAS-CI calculations with Dirac-Coulomb Hamiltonian, $n=1-3$ and $4s4p$ shells were kept frozen~\cite{Ramanantoanina:22}, QED~\cite{Lowe:13} and Breit corrections were estimated at the MCDF level; \\
$^{e)}$ configuration interaction calculations with DCB Hamiltonian using numerical SCF spinors~\cite{Dzuba:14}, QED is included via the model potential of Flambaum et al.~\cite{Flambaum:05}.
\end{table}
It can be seen from Table~\ref{tab:lu} that the final accuracy of the RPP calculation including triples correction (see the ``SD+$\Delta_T$'' column) is considerably worse than for the previously discussed cases of Tl$^+$ and Ra. For the upper half of the energy interval considered, the account of triples greatly reduces the errors, at the same time slightly deteriorating the results for the lowest electronic states. However, the accuracy becomes more balanced. A similar pattern can be observed if we pass from the minimum possible active space of $6s6p5d$ spinors to the extended active space including the next shells ($7s7p6d5f$). In this case, some part of triple excitations is effectively accounted for, but the most important ones involving the hole in the $4f$ shell are missed. This leads to an imbalanced treatment of triples and larger errors when compared to the additive scheme based on the genuine FS-RCCSDT approximation. Furthermore, in such strongly correlated systems like Lu$^+$ quadruple excitations can also play an important role~\cite{Oleynichenko:CCSDT:20}; unfortunately, the FS-RCCSDTQ model which is able to evaluate such corrections is extremely computationally demanding and is not available to date.

The next feature to be noted is the fact that the error for the given state is nearly proportional to the number of $d$-electrons in its leading configuration. For the $6s6p$ states the errors do not exceed those obtained for the Tl$^+$ case and the accuracy can be regarded as quite satisfactory. This may be related to a significant effect of the lack of basis functions with high angular momenta (at least $k$-functions) presently unavailable in RPP calculations. To estimate the contributions from high-$l$ basis functions we performed the series of all-electron FS-RCCSD calculations within the molecular mean field approximation~\cite{Sikkema:09} to the Dirac-Coulomb-Gaunt Hamiltonian. For the value of the $k$-function exponent equal to 1.2 contributions to excitation energies were found to be of order $-$3 cm$^{-1}$, $-$80 cm$^{-1}$, $-$85 cm$^{-1}$ and up to $-$160 cm$^{-1}$ for $6s6p$, $6p5d$, $6s5d$ and $5d^2$ states, respectively. Such large corrections for $d$-states clearly indicate the necessity of inclusion of at least $k$-functions into the basis set. The overall error arising from the lack of higher harmonics is expected to be at least twice as much, up to $-$250 cm$^{-1}$ for the $5d^2$ states. This effect is 
expected to be even more pronounced if we included $k$-functions into the basis at the FS-RCCSDT calculation step. One can argue that the presence of the closed $4f$ shell leads to strong angular correlations which have to be thoroughly accounted for to achieve accuracy of order $\sim$ 100 cm$^{-1}$ for $d$-states. This circumstance can be of crucial importance for highly accurate predictions of atomic energy levels of lutetium's heavier homolog, lawrencium~\cite{Borschevsky:07,Kahl:19,Kahl:21,Ramanantoanina:22}.

The tiny-core RPP used for Lu replaces its 28 inner core electrons. To estimate contributions of the core-valence correlation with these electrons we have also performed all-electron Dirac-Coulomb-Gaunt FS-RCCSD calculations explicitly treating the $3s3p3d$ shells. Contributions to Lu$^{+}$ excitation energies did not exceed +20 cm$^{-1}$ for the vast majority of states except the last five ones, for which the contributions from the $n=3$ shell electrons reached $\sim$30-50 cm$^{-1}$. This completely legitimizes the use of the tiny-core 28-electron RPP for Lu.

Finally, it should be pointed out that the best results for Lu$^{+}$ excitation energies obtained within state-of-the-art relativistic atomic calculations~\cite{Kahl:19,Kahl:21,Dzuba:14} are not fundamentally more accurate than those obtained in the present work within the computationally much more cheap RPP-QED model for the Hamiltonian. In the last two columns of Table~\ref{tab:lu} we compare QED corrections obtained in the present work and calculated within the completely different approach~\cite{Kahl:19} (atomic four-component CI+MBPT, model QED potential of Flambaum et al.~\cite{Flambaum:05}). The excellent agreement between them again proves the correctness of the RPP-QED approach.

\section{Concluding remarks}
\label{sec:conclusion}

In the present work, a simple procedure to incorporate one-loop quantum electrodynamic (QED) corrections into the generalized shape-consistent relativistic pseudopotential model is proposed. The only required modification of the conventional procedure of GRPP construction consists in adding the model Lamb shift potential to the Dirac-Coulomb-Breit many-electron Hamiltonian 
defining the reference all-electron atomic SCF problem. This paves the way towards routine inclusion of QED in electronic structure calculation of molecules, which is especially interesting because of 
the importance of such contributions for highly precise calculation of molecular properties recently proven in~\cite{Skripnikov:QED:21,Sunaga:QED:22}. Pilot applications of the new model to calculations of excitation energies of two-valence-electron atomic systems with increasing complexity (Ra, Tl$^+$, Lu$^+$) allows one to draw the following conclusions about the accuracy of the approach developed:
\begin{itemize}
\item[--] for the chosen 6-7 row elements, the deviations of SCF estimates of electronic transition energies by the GRPP approximation from their counterparts obtained with DCB Hamiltonian combined with QED model potential are by an order of magnitude smaller than those arising from the neglect of QED corrections;
\item[--] the accuracy of the constructed GRPP model in correlation calculations is mainly restricted by the neglect of correlations involving inner-core electrons and thus depends crucially on the sizes of excluded inner cores. The results of correlation calculations on Tl$^+$ and Lu$^+$ with GRPP of 28-electron cores (main quantum number $n \le 3$) allow 
to expect that this model is sufficiently accurate to simulate excitation energies of 6$^{\rm th}$ row atoms and their compounds with an accuracy better than 10$^2$ cm$^{-1}$. As for the case of the Ra atom, the possibility of further improvement of accuracy through reducing the inner core size from 60 to 28 electrons deserves additional studies;
\item[--] even for very small inner cores, the calculations with the GRPP Hamiltonian remain much less expensive than those with 4-component ones, at the same time ensuring a comparable accuracy of modeling low-energy electronic excitations as the most advanced 4-component approximation. Due to the proper account of electron-electron interactions beyond the Coulomb 
term, the ``tiny-core'' GRPP models clearly outperform in accuracy the Dirac--Coulomb Hamiltonian. Thus the GRPPs define the relativistic electronic structure model with an excellent cost 
to accuracy ratio.
\end{itemize}

The proposed reformulation of the intermediate-Hamiltonian FS-RCC method 
provides an efficient tool for studies on excited states of heavy element atoms and molecules of their compounds, offering the possibility to systematically improve the results through model space extension and accounting for higher excitations in the cluster operator. This long-awaited feature will be indispensable in forthcoming molecular applications using the presented pseudopotentials.

The detailed analysis of problems 
that emerged during the modeling of Lu$^+$ electronic states has clearly demonstrated the 
issue of the shortage of high-angular-momentum basis functions when modeling systems with the 
valence or even high-energy filled $f$-shell. Now it is quite obvious that further improvements in the precision of calculations on heavy-element compounds with GRPPs using a conventional basis set expansions of one-electron spinors will require 
the development of efficient codes for evaluating the GRPP matrix elements with high-angular-momentum basis functions. Such a code developed by our group will be presented in 
a forthcoming publication.



\section*{Acknowledgements}

We are indebted to Anatoly V. Titov for helpful discussions and critical reading of the manuscript.

\section*{Funding Information}

The work of AZ, NSM and AVO at NRC ``Kurchatov Institute'' - PNPI on the RPP generation for the Tl and Lu atoms accounting for the QED effects, development and implementation of the incomplete main model space version of intermediate Hamiltonian FS-RCC with a single-sector shift, and pilot calculations was supported by the Russian Science Foundation (Grant No. 20-13-00225). The contribution of EE concerning the formulation of multisector generalization of the shift technique was partially financed by the Ministry of Science and Higher Education of the Russian Federation within Grant No. 075-10-2020-117.

\section*{Research Resources}

Calculations have been carried out using computing resources of the federal collective usage center Complex for Simulation and Data Processing for Mega-science Facilities at National Research Centre ``Kurchatov Institute'', http://ckp.nrcki.ru/.

\section*{Conflict of interest}

The authors declare no conflict of interest.

\selectlanguage{english}
\FloatBarrier

\bibliographystyle{achemso.bst}
\bibliography{qedrpp.bib}

\end{document}